\documentclass[12pt]{article}
\usepackage[dvips]{graphicx}
\usepackage{epsfig,amsmath}

\begin{document}
\thispagestyle{empty}
\begin{center} 
{\bf RECENT PROGRESS IN THE STATISTICAL APPROACH OF PARTON DISTRIBUTIONS \footnote{Invited talk at DIFFRACTION 2010, Otranto, Italy, September 10 - 15, 2010, to appear in the AIP Conference Proceedings}}\\

\vskip 1.4cm
{\bf Jacques Soffer}
\vskip 0.3cm
{\it Physics Department, Temple University},\\ 
{\it Philadelphia, PA 19122-6082, USA}\\
\end{center}
\vskip 1.5cm

{\bf Abstract}
We recall the physical features of the parton distributions in the quantum statistical approach of the nucleon. Some
 predictions from a next-to-leading order QCD analysis are compared to recent experimental results. We also consider their extension to 
 include their transverse momentum dependence.\\


\vskip 2.0cm

%%%%%%%%%%%%%%%%%%%%%%%%%%%%%%%%%%%%%%%%%%%%
%% MAINMATTER
%%%%%%%%%%%%%%%%%%%%%%%%%%%%%%%%%%%%%%%%%%%%
Let us first review some of the basic features of the statistical approach, as oppose
to the standard polynomial type
parametrizations of the parton distribution functions (PDF), based on Regge theory at low $x$ and counting
rules at large $x$.
The fermion distributions are given by the sum of two terms \cite{bbs1},
a quasi Fermi-Dirac function and a helicity independent diffractive
contribution equal for all light quarks:
\begin{equation}
xq^h(x,Q^2_0)=
\frac{AX^h_{0q}x^b}{\exp [(x-X^h_{0q})/\bar{x}]+1}+
\frac{\tilde{A}x^{\tilde{b}}}{\exp(x/\bar{x})+1}~,
\label{eq1}
\end{equation}
\begin{equation}
x\bar{q}^h(x,Q^2_0)=
\frac{{\bar A}(X^{-h}_{0q})^{-1}x^{2b}}{\exp [(x+X^{-h}_{0q})/\bar{x}]+1}+
\frac{\tilde{A}x^{\tilde{b}}}{\exp(x/\bar{x})+1}~,
\label{eq2}
\end{equation}
at the input energy scale $Q_0^2=4 \mbox{GeV}^2$.
Notice the change of sign of the potentials
and helicity for the antiquarks.
The parameter $\bar{x}$ plays the role of a {\it universal temperature}
and $X^{\pm}_{0q}$ are the two {\it thermodynamical potentials} of the quark
$q$, with helicity $h=\pm$. It is important to remark that the diffractive contribution 
occurs in the unpolarized distributions $q(x)= q_{+}(x)+q_{-}(x)$, but it is absent in the valence $q_v(x)= q(x) - \bar {q}(x)$ and in the helicity
distributions $\Delta q(x) = q_{+}(x)-q_{-}(x)$ (similarly for antiquarks).
The {\it eight} free parameters\footnote{$A=1.74938$ and $\bar{A}~=1.90801$ are
fixed by the following normalization conditions $u-\bar{u}=2$, $d-\bar{d}=1$.}
in Eqs.~(\ref{eq1},\ref{eq2}) were
determined at the input scale from the comparison with a selected set of
very precise unpolarized and polarized Deep Inelastic Scattering (DIS) data \cite{bbs1}. They have the
following values
\begin{equation}
\bar{x}=0.09907,~ b=0.40962,~\tilde{b}=-0.25347,~\tilde{A}=0.08318,
\label{eq3}
\end{equation}
\begin{equation}
X^+_{0u}=0.46128,~X^-_{0u}=0.29766,~X^-_{0d}=0.30174,~X^+_{0d}=0.22775~.
\label{eq4}
\end{equation}
For the gluons we consider the black-body inspired expression
\begin{equation}
xG(x,Q^2_0)=
\frac{A_Gx^{b_G}}{\exp(x/\bar{x})-1}~,
\label{eq5}
\end{equation}
a quasi Bose-Einstein function, with $b_G=0.90$, the only free parameter
\footnote{In Ref.~\cite{bbs1} we were assuming that, for very small $x$,
$xG(x,Q^2_0)$ has the same behavior as $x\bar q(x,Q^2_0)$, so we took $b_G = 1
+ \tilde b$. However this choice leads to a too much rapid rise of the gluon
distribution, compared to its recent  determination from HERA data, which
requires $b_G=0.90$.}, since $A_G=20.53$ is determined by the momentum sum
rule.
 We also assume that, at the input energy scale, the polarized gluon,
distribution vanishes, so
\begin{equation}
x\Delta G(x,Q^2_0)=0~.
\label{eq6}
\end{equation}
For the strange quark distributions, the simple choice made in Ref. \cite{bbs1}
was greatly improved in Ref. \cite{bbs2}. More recently, new tests against experimental (unpolarized and
polarized) data turned out to be very satisfactory, in particular in hadronic
reactions, as reported in Refs.~\cite{bbs3,bbs4}.\\
For illustration, we will just give two recent results, directly related to the determination of the quark distributions from unpolarized
and polarized DIS. First, we display on Fig.~1($\it{Left}$), the resulting unpolarized statistical PDF versus $x$ at $Q^2$=10 $\mbox{GeV}^2$, where $xu_v$ is the $u$-quark valence, $xd_v$ the $d$-quark valence, with their characteristic maximum around $x=0.3$, $xG$ the gluon and $xS$
stands for twice the total antiquark contributions, $\it i.e.$ $xS(x)=2x(\bar {u}(x)+ \bar {d}(x) + \bar {s}(x))+ \bar {c}(x))$. Note that $xG$ and $xS$ are downscaled by a factor 0.05. They can be compared with the parton distributions as determined by the H1PDF 2009 QCD NLO fit, shown also in Fig.~1($\it{Right}$), and the agreement is rather good. The results are based on recent $ep$ collider data from HERA, combined with previously published data and the accuracy is typically
in the range of 1.3 - 2 $\%$.\\
Concerning the light antiquark helicity distributions, the statistical 
approach
imposes a strong relationship to the corresponding quark helicity
distributions. In particular, it predicts $\Delta \bar u(x)>0$ and $\Delta \bar
d(x)<0$, with almost the same magnitude, in contrast with the
simplifying assumption $\Delta \bar u(x)=\Delta \bar d(x)$, often adopted in
the literature. The COMPASS experiment
at CERN has measured the valence quark helicity distributions, defined as
$\Delta q_v(x)= \Delta q(x)-\Delta \bar q(x)$. These recent results displayed
in Fig.~2 are compared to our prediction and the data give
$\Delta \bar u(x) + \Delta \bar d(x) \simeq 0$, which implies either small or
opposite values for $\Delta \bar u(x)$ and $\Delta \bar d(x)$. Indeed $\Delta
\bar u(x)>0$ and $\Delta \bar d(x)<0$, predicted by
the statistical approach \cite{bbs1}, lead to a non negligible
 positive contribution of the sea to the Bjorken sum rule, an interesting consequence.\\
We now turn to another important aspect of the statistical PDF and very briefly discuss
a new version of the extension to the transverse momentum dependence (TMD).
In Eqs.~(\ref{eq1},\ref{eq2}) the multiplicative factors $X^{h}_{0q}$ and
$(X^{-h}_{0q})^{-1}$ in
the numerators of the non-diffractive parts of $q$'s and $\bar{q}$'s
distributions, imply a modification
of the quantum statistical form, we were led to propose in order to agree with
experimental data. The presence of these multiplicative factors was justified
in our earlier attempt to generate the TMD \cite{bbs5}, but it was not properly done and 
a considerable improvement was achieved recently \cite{bbs6}. We have introduced some thermodynamical
potentials $Y^h_{0q}$, associated to the quark transverse momentum $k_T$, and
related to $X^{h}_{0q}$ by the simple relation
$\mbox{ln}(1+\exp[Y^h_{0q}])=kX^h_{0q}$. We were led to choose $k=3.05$ and this method involves another parameter
$\mu^2$, which plays the role of the temperature for the transverse degrees of
freedom and whose value was determined by the transverse energy sum rule.
\begin{figure}[htb]
  %\hspace*{-10mm}
  \begin{minipage}{7.0cm}
  \epsfig{figure=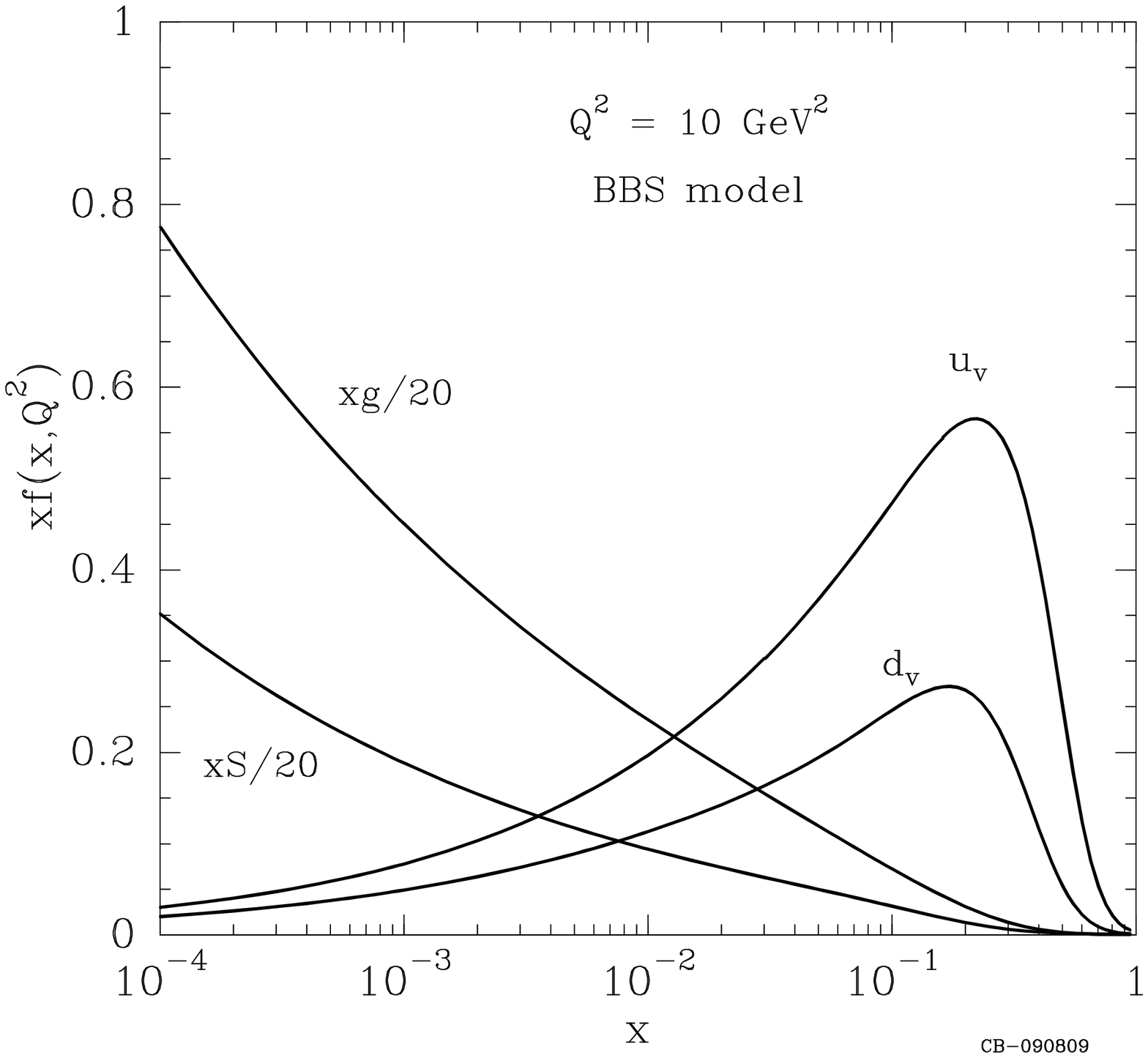,width=7.4cm}
  \end{minipage}
  %\hspace*{-2mm}
    \begin{minipage}{7.0cm}
  \epsfig{figure=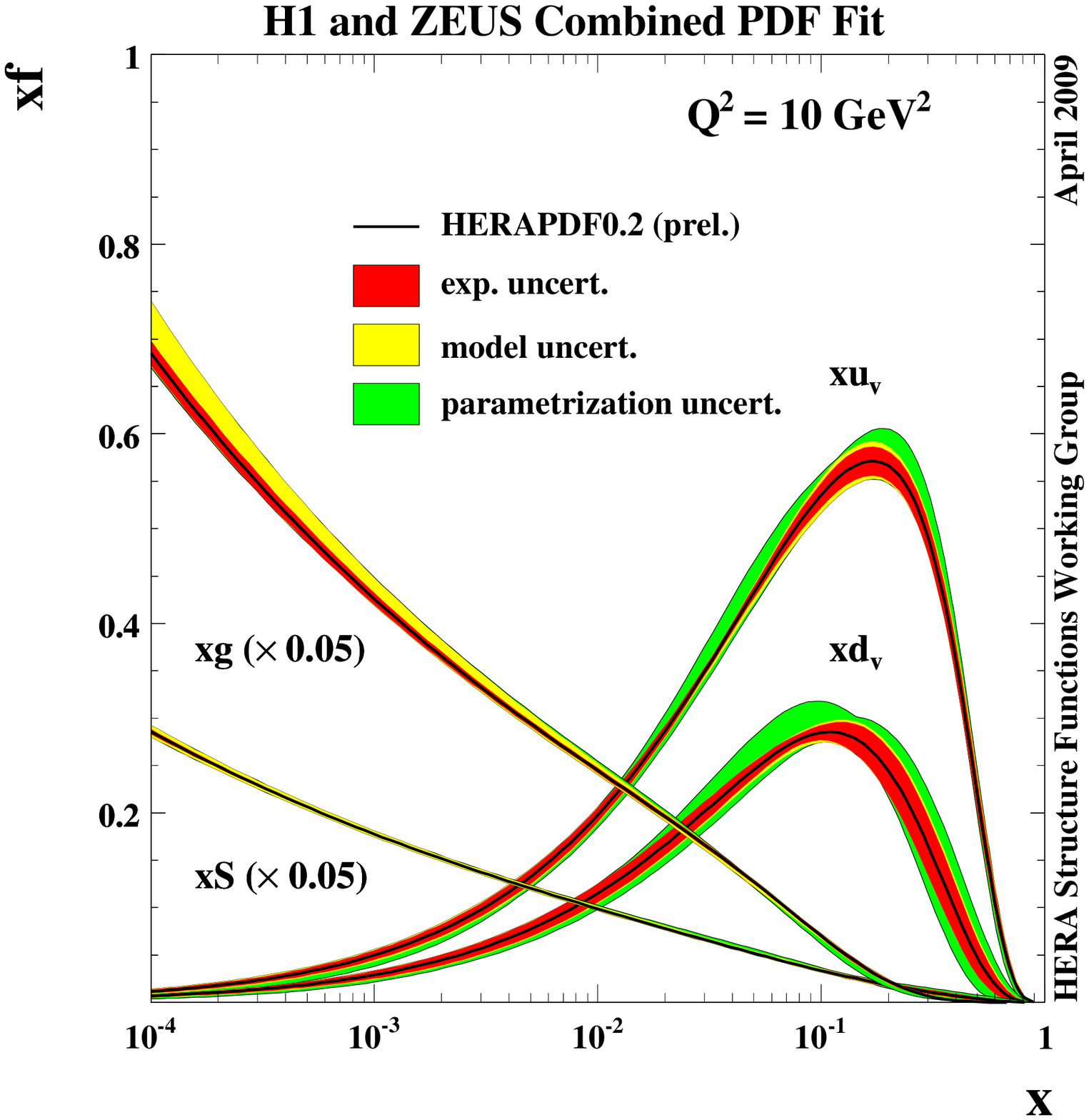,width=6.8cm}
  \vspace*{-10mm}
  \end{minipage}\\
\caption{
{\it Left} : BBS predictions for various statistical unpolarized parton distributions versus $x$ at $Q^2=10\mbox{GeV}^2$. {\it Right} : Parton distributions at $Q^2=10\mbox{GeV}^2$, as determined by the H1PDF fit, with different uncertainties (Taken from Ref. \cite{aaron}).}
\label{fi:fig1}
\end{figure}
\begin{figure}[htb]
  \hspace*{+25mm}
  \epsfig{angle=-90,figure=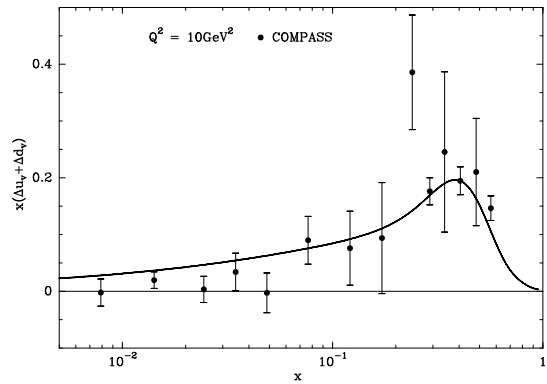,width=8.0cm}
  \vspace*{-3mm}
\caption{ The valence quark helicity distributions versus $x$ and evolved at $Q^2=10\mbox{GeV}^2$. The solid curve is the BBS prediction
of the statistical approach and the data come from Ref. \cite{compass}.}
\label{fi:fig2}
\end{figure}
\clearpage
 We have calculated the $p_T$ dependence of semiinclusive DIS cross sections and double
longitudinal-spin asymmetries, taking into account the effects of the
Melosh-Wigner rotation, for $\pi^{\pm}$ production by using this
 set of TMD statistical parton distributions and another set coming from the relativistic
covariant approach \cite{zav}. Both sets do not satisfy the usual factorization
assumption of the dependence in $x$ and $k_T$ and they lead to different results, which can be compared
to recent experimental data from CLAS at JLab, as shown on Fig.~3.
\begin{figure}[htb]
  \begin{minipage}{7.0cm}
  \hspace*{+6mm}
  \epsfig{figure=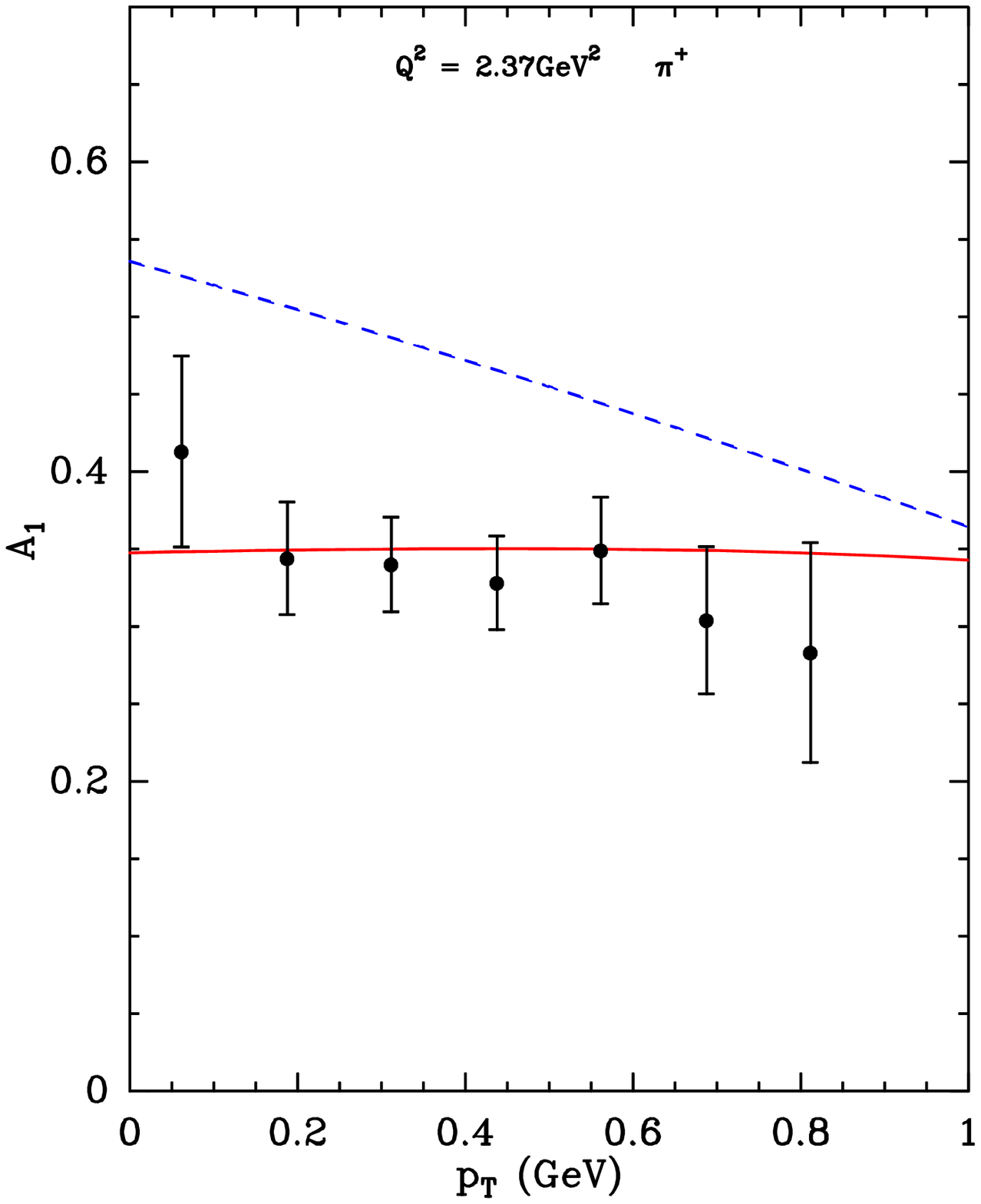,width=7.5cm}
  \end{minipage}
    \hspace*{-4mm}
    \begin{minipage}{7.0cm}
  \epsfig{figure=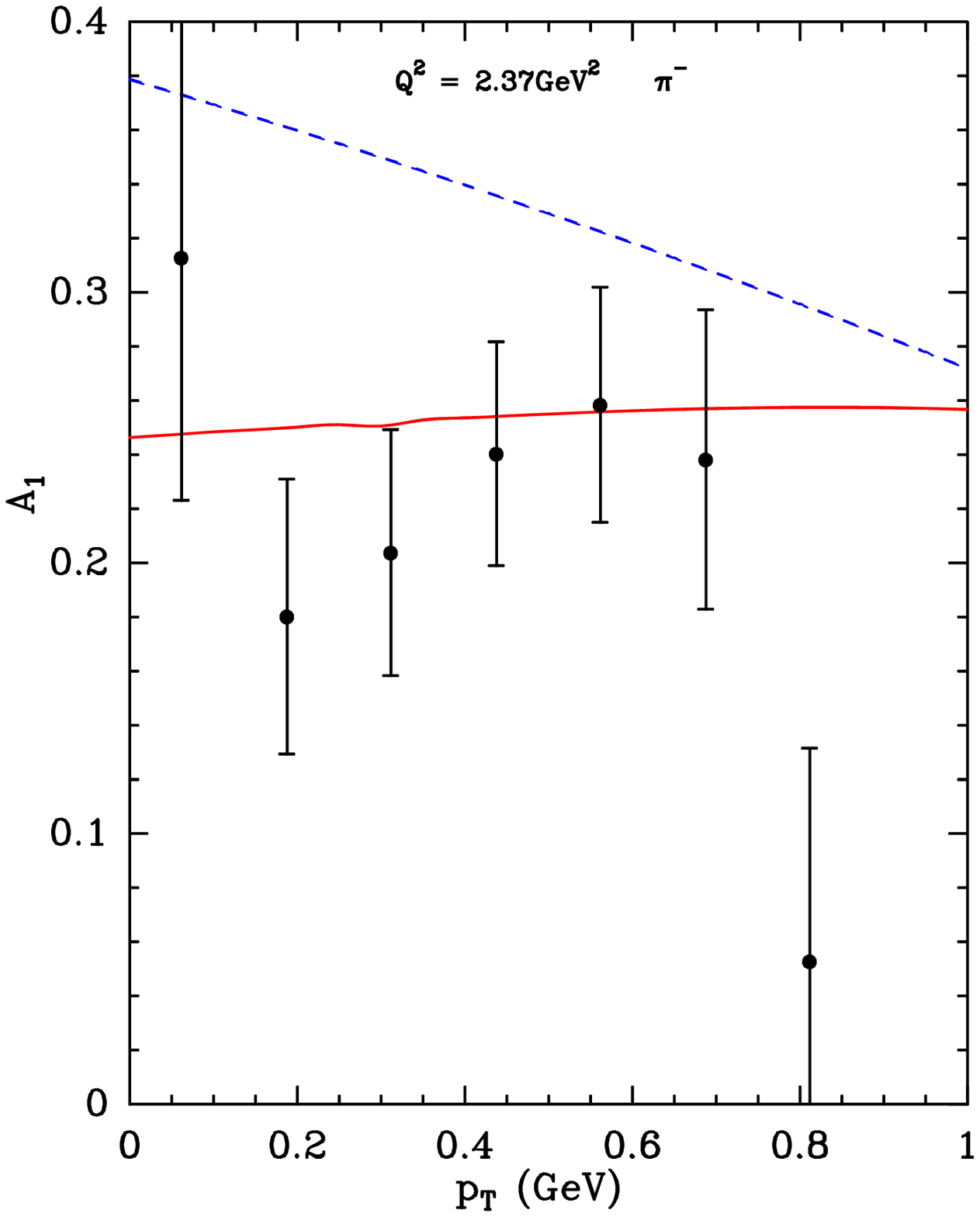,width=7.5cm}
  \vspace*{-3mm}
    \end{minipage}\\
\caption{The double longitudinal-spin asymmetry $A_1$ for $\pi^+$ ($\it left$) and
$\pi^-$ ($\it right$) production on a proton target, versus the $\pi$ momentum $p_T$, compared
 to the JLab data Ref.~\cite{clas2}. The solid lines are the
results from the TMD statistical distributions \cite{bbs6} and the dashed lines correspond to
the relativistic covariant distributions \cite{zav}.}
\label{fi:fig3}
%\vspace*{-2.5ex}
\end{figure}

%%%%%%%%%%%%%%%%%%%%%%%%%%%%%%%%%%%%%%%%%%%%


\begin{thebibliography}{99}

\bibitem{bbs1} C. Bourrely, F. Buccella and J. Soffer,
\emph{Eur. Phys. J. C} {\bf 23}, 487 (2002).

\bibitem{bbs2} C. Bourrely, F. Buccella and J. Soffer,
\emph{Phys. Lett. B} {\bf 648}, 39 (2007).

\bibitem{bbs3}  C. Bourrely, F. Buccella and J. Soffer,
\emph{Mod. Phys. Lett. A} {\bf 18}, 771 (2003).

\bibitem{bbs4} C. Bourrely, F. Buccella and J. Soffer,
\emph{Eur. Phys. J. C} {\bf 41}, 327 (2005).

\bibitem{aaron} F. D. Aaron {\it et al.}, [H1 Collaboration], 
\emph{Eur. Phys. J. C} {\bf 64}, 561 (2009).

\bibitem{compass} M. Alekseev {\it et al.} [COMPASS Collaboration], \emph{Phys. Lett. B} {\bf 660}, 458 (2008).

\bibitem{bbs5} C. Bourrely, F. Buccella and J. Soffer,
\emph{Mod. Phys. Lett. A} {\bf 21}, 143 (2006).

\bibitem{bbs6} C. Bourrely, F. Buccella and J. Soffer, arXiv:1008.5322v1 [hep-ph].

\bibitem{zav} P. Zavada,
\emph{Eur. Phys. J. C} {\bf 52}, 121 (2007) and references therein.
 A.V. Efremov, P. Schweitzer, O.V. Teryaev and P. Zavada,
Proceedings of XIII Workshop on High Energy Spin Physics DSPIN-09, Dubna, 
Russia, September 1-5, 2009.
 arXiv:0912.3380v3 [hep-ph] and references therein. See also arXiv:1008.3827v1 
[hep-ph].


\bibitem{clas2} H. Avakian $\it{et~al.}$ (CLAS), arXiv:1003.4549v1 [hep-ex].




\end{thebibliography}
\end{document}